\documentclass[aps,prl,onecolumn,preprintnumbers,groupedaddress]{revtex4}
\usepackage{epsfig,graphics,amssymb,amsmath,subeqnarray}

\newcommand \bnabla{\boldsymbol{\nabla}}
\newcommand \bu{\mathbf{u}}
\newcommand \buc{\mathbf{\tilde{u}}}
\newcommand \UD{U_\Delta}
\newcommand \UDone{U_{\Delta 1}}
\newcommand \UDtwo{U_{\Delta 2}}
\newcommand \btau{\boldsymbol{\tau}}
\newcommand \btauc{\tilde{\boldsymbol{\tau}}}
\newcommand \bgam{\dot{\boldsymbol{\gamma}}}
\newcommand \bgamc{\tilde{\dot{\boldsymbol{\gamma}}}}
\newcommand \bsigma{\boldsymbol{\sigma}}
\newcommand \psic{\tilde{\psi}}
\newcommand \Deone{\text{De}_1}
\newcommand \Detwo{\text{De}_2}

\newcommand \T{\text{T}}
\def\d{{\rm d}}

\begin{document}

\title{Two-dimensional flagellar synchronization in  viscoelastic fluids}
\author{Gwynn J Elfring}
\author{On Shun Pak}
\author{Eric Lauga}
\email{elauga@ucsd.edu}
\affiliation{
Department of Mechanical and Aerospace Engineering, 
University of California San Diego,
9500 Gilman Drive, La Jolla CA 92093-0411, USA.}
\date{\today}
\begin{abstract}
Experimental studies have demonstrated that  spermatozoa  synchronize their flagella when swimming in close proximity. In a Newtonian fluid, it was shown theoretically that such synchronization arises passively due to hydrodynamic forces between the two swimmers if their waveforms exhibit a front-back geometrical asymmetry. Motivated by the fact that most biological fluids possess a polymeric microstructure, we  address here synchronization in a viscoelastic fluid analytically. Using a two-dimensional infinite sheet model we show that the presence of polymeric stresses removes the geometrical asymmetry constraint, and therefore even symmetric swimmers synchronize. Such synchronization occurs on asymptotically faster time scales than in a Newtonian fluid, and  the swimmers are seen to be driven into a stable in-phase conformation minimizing the energy dissipated in the surrounding fluid.

\end{abstract}


\maketitle

\section{Introduction}

Swimming microorganisms are found everywhere in nature, from bacteria moving towards nutrients, to spermatozoa migrating towards an ovum. Given their size, they swim in a realm where viscous stresses predominate and inertia is negligible. This fact often necessitates strategies for locomotion which are vastly different from those of organisms  in the macroscopic world \cite[]{purcell}. Much progress has been made in the physical and hydrodynamic analysis of microscopic swimming \cite[]{lighthill,brennan,lauga2}, yet  many marked phenomena are still of current interest.  One such phenomenon is the experimentally-observed synchronization of spermatozoa flagella (slender filaments deformed in a wavelike fashion and propelling the cells forward) when  swimming in close proximity \cite[]{yang,woolley}. This synchronization has been observed to prompt an increase in the swimming speed of the co-swimming cells which could thereby provide a competitive, and hence perhaps evolutionary, advantage \cite[]{woolley}. The theoretical analysis of the synchronization of swimming microorganisms in Newtonian fluids dates back to the work of Taylor who, assuming two  infinite and parallel two-dimensional sheets, showed that synchronous beating minimizes the viscous dissipation between the model cells \cite[]{taylor1}. This synchronization was subsequently demonstrated computationally for both infinite \cite[]{fauci1}, and finite two-dimensional models \cite[]{fauci2}. Recently, it was shown  that phase-locking arises purely passively due to the  fluid forces between the two swimmers, and requires  a front-back asymmetry in the flagellar geometry of the cells without which no synchronization can occur \cite[]{elfring}.

In all previously-studied situations, synchronization was addressed in the case of a  Newtonian fluid. However, most biological fluids involved, for example,  in mammalian reproduction  are non-Newtonian. As mammalian spermatozoa make their journey through the female reproductive tract they encounter several complex fluids, including glycoprotein-based cervical mucus in the cervix, mucosal epithelium inside the fallopian tubes, and actin-based viscoelastic gel outside the ovum \cite[]{suarez06,dunn76}. In a viscoelastic fluid, kinematic reversibility, restated in Purcell's scallop theorem \cite[]{purcell}, breaks down due to the presence of normal stresses and shear-dependent material functions, fundamentally altering the  governing flow physics \cite[]{lauga3}. The waveform, structure, and swimming path of spermatozoa have been experimentally observed to be modified in viscoelastic fluids \cite[]{fauci3}. Locomotion in complex fluids has been studied analytically \cite[]{chaudhury,sturges,fulford}, and it has been shown that microorganisms which propel themselves by propagating waves along their flagella have a lower swimming speed in a viscoelastic fluid than in a Newtonian fluid \cite[]{lauga1, fu}.

In this paper we study the passive  synchronization of two flagellated cells in a viscoelastic (Oldroyd-B) fluid. Using Taylor's infinite two-dimensional sheet model, we show that not only does phase locking arise in a viscoelastic fluid, but also that it does not require the front-back geometrical asymmetry that must exist for such a model to display synchronization in a Newtonian fluid. We demonstrate that the  system evolves to a single stable fixed point at the in-phase conformation, which is also the conformation that yields minimal energy dissipation. In addition,  we show that the evolution to a phase-locked state occurs on asymptotically faster time scales than in a Newtonian fluid.

\section{Setup}
\begin{figure}
\centerline{\includegraphics[width=.85\textwidth]{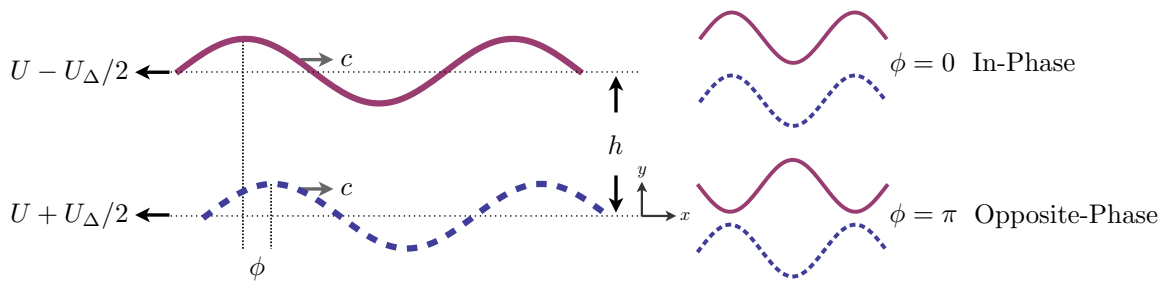}}
\caption{Model system consisting of two infinite sinusoidal sheets passing waves at speed $c$ and thereby swimming at speed $U\pm\UD/2$ in the opposite direction. The difference in phase $\phi$ incurs a relative velocity $\UD(\phi,h)$ between the two sheets denoted positive when the top sheet  swims to the right relative to the bottom one. The average separation distance is denoted $h$.}
\label{system}
\end{figure}

Our system, shown in Fig.~\ref{system}, consists of two parallel infinite two-dimensional sheets, separated by a mean distance $h$. Both sheets propagate sinusoidal waves of transverse displacement of amplitude $a$ at speed $c=\omega/k$, where $\omega$ is the wave frequency and $k$ is the wavenumber, but have an initial phase difference $\phi_0$. By passing these waves, the sheets propel themselves  in the direction opposite to the wave speed \cite[]{taylor1}.  The sheets are also permitted to move relative to each other with an unknown velocity $\UD$, denoted positive when the top sheet  (\#2) swims in the positive $x$ direction relative to the bottom one (\#1). The positions of the sheets, in their swimming frames, are thereby given by $y_1=a\sin(kx-\omega t -\phi_0/2+\int_{0}^{t} k \UD(t')dt'/2)$ and $y_2=h+a\sin(kx-\omega t + \phi_0/2-\int_{0}^{t} k \UD(t')dt'/2)$.

We use the following dimensionless variables $\hat{x}^*=xk$, $t^*=t\omega$, $\mathbf{u}^*=\mathbf{u}/c$, $\UD^*=\UD/c$. The amplitude of the waves is non-dimensionalized by the wavenumber, $\epsilon=ak$. For convenience we let $x^*=\hat{x}^*-t^*$ and $\phi=\phi_0-\int_{0}^{t^*} \UD^*(t')dt'$ which is the instantaneous phase difference between the two sheets. The position of the sheets in dimensionless form is thus given by $y_1^*=\epsilon \sin(x^*-\phi/2)$, and $y_2^*=h^*+\epsilon \sin(x^*+\phi/2)$, and  the phase evolves in time according  to $\dot{\phi}=-\UD^*$. We refer to the $\phi=0$ conformation as in-phase, and the $\phi=\pi$ conformation as opposite-phase. The system is $2\pi$ periodic and $\phi$ is defined from $-\pi$ to $\pi$. We now drop the ($^*$) notation and refer only below to dimensionless variables.

Since this problem is two dimensional, we introduce the streamfunction $\psi(x,y,t)$ where the components of the velocity field are  $\bu=[u,v]^\T=[\partial\psi/\partial y,-\partial\psi/\partial x]^\T$, and the incompressibility condition is always satisfied. The boundary conditions become then
\begin{eqnarray}
\bnabla\psi|_{y= y_1} \ &=& \bnabla\Big[-\UD y/2+\epsilon\sin(x-\phi/2)\Big]\Big|_{y= y_1},\\
\bnabla\psi|_{y= y_2} \ &=& \bnabla\Big[\UD y/2+\epsilon\sin(x+\phi/2)\Big]\Big|_{y= y_2}.
\end{eqnarray}

Mechanical equilibrium for a low-Reynolds number flow in a viscoelastic fluid is given by $\boldsymbol{\nabla}p=\bnabla\cdot\boldsymbol{\tau}$, where $p$ is the pressure and $\btau$ is the deviatoric part of the stress tensor. To relate the stress to the strain-rate, we use in this paper the simplest  
polymeric constitutive equation derived from a microscopic theory, namely the 
Oldroyd-B model \cite[]{oldroyd,bird1}  given by
\begin{equation}
\boldsymbol{\tau}+\Deone \stackrel{\triangledown}{\btau}=\bgam+\Detwo\stackrel{\triangledown}{\bgam},
\label{oldroyd}
\end{equation}
where $\bgam=\bnabla\bu+\bnabla\bu^\T$ is the strain-rate tensor; the upper convected derivative is defined for a general tensor $\mathbf{A}$ as $\stackrel{\triangledown}{\mathbf{A}}=\partial \mathbf{A}/\partial t +\bu\cdot\bnabla\mathbf{A}-(\bnabla\bu^{\T}\cdot\mathbf{A}+\mathbf{A}\cdot\bnabla\bu)$. We have defined two Deborah numbers, $\Deone=\lambda_1 \omega$, where $\lambda_1$ is the relaxation time of the polymer, and $\Detwo=\lambda_2 \omega$, where $\lambda_2=\lambda_1\eta_s/(\eta_s+\eta_p)$ is the retardation time of the polymer and $\eta_s$, $\eta_p$ refer to the contribution to the viscosity from the solvent and polymer respectively \cite[]{bird1}. Rheological studies have shown the relaxation time of cervical mucus to be between 1 and 10 seconds \cite[]{tam80}. Given that the flagella of spermatozoa typically beat at a frequency between 20 and 50 Hz \cite[]{brennan} we might expect a range of Deborah numbers $\Deone=10^2-10^3$, and in most practical instances $\Deone\gg\Detwo$ \cite[]{bird1}.

In the spirit of Taylor's seminal study, we look to solve this system in a small amplitude limit, $\epsilon \ll 1$, with a perturbation expansion in powers of $\epsilon$ for stress $\btau=\epsilon\btau_1+\epsilon^2\btau_2+...$, strain-rate $\bgam=\epsilon\bgam_1+\epsilon^2\bgam_2+...$, streamfunction $\psi=\epsilon\psi_1+\epsilon^2\psi_2+...$ and the relative velocity between the sheets $\UD=\epsilon\UDone+\epsilon^2\UDtwo+...$.

\section{Analysis}

\subsection{First-order solution}
The leading order component of \eqref{oldroyd} is
\begin{equation}
\btau_1+\Deone\frac{\partial \btau_1}{\partial t}=\bgam_1+\Detwo\frac{\partial \bgam_1}{\partial t}
\label{oldroyd1}\cdot
\end{equation}
Taking the divergence and the curl of \eqref{oldroyd1} we get the governing equation for the first-order streamfunction
\begin{equation}
\left(1+\Detwo\frac{\partial}{\partial t}\right)\nabla^4\psi_1=0.
\label{biharmonic1}
\end{equation}
With the first order boundary conditions
\begin{eqnarray}
\bnabla\psi_1|_{y= 0} \ &=& \bnabla\left[(-\UDone y/2+\sin(x-\phi/2)\right]|_{y= 0},\\
\bnabla\psi_1|_{y= h} \ &=& \bnabla\left[\UDone y/2+\sin(x+\phi/2)\right]|_{y= h},
\end{eqnarray}
the solution can be shown to be
\begin{equation}\label{stream1}
\psi_1=a_0(y)+a_1(y)\cos(x)+b_1(y)\sin(x),
\end{equation}
where
\begin{eqnarray}
a_0(y) &=& \text{C}_1 y^2 \left( y-\frac{3}{2} h\right)+\frac{1}{2} \UDone y \left(\frac{y}{h}-1\right), \label{g3}\\
a_1(y) &=& \frac{\sin\frac{\phi }{2}}{h-\sinh(h)} \Big(y \cosh(h-y)-(h-y) \cosh y +\sinh(h-y)-\sinh y\Big), \\
b_1(y) &=& \frac{\cos\frac{\phi }{2}}{h+\sinh (h)} \Big(y \cosh(h-y)+(h-y) \cosh y +\sinh(h-y)+\sinh y\Big).
\end{eqnarray}

To determine the unknown constant $\text{C}_1$ we resort to dynamical considerations. For simplicity we resolve the streamfunction into its complex Fourier components in the wave variable $x=\hat{x}-t$ giving $\psi_1=\Re\left[\psic_1^{(0)}+\psic_1^{(1)}\right]$, where $\Re[...]$ denotes the real part and $\tilde{\psi_1}^{(0)}=a_0(y)$ and $\tilde{\psi}_1^{(1)}=(a_1(y)+ ib_1(y))e^{-ix}$. The strain-rate tensor, $\bgam_1=\bnabla{\bu}_1+\bnabla{\bu}_1^\T$, can then be  obtained using \eqref{stream1}. Exploiting  \eqref{oldroyd1}, we see that the first-order stress tensor is given by
\begin{equation}
\btau_1=\Re \left[\bgamc_1^{(0)}+\frac{1+i\Detwo}{1+i\Deone}\bgamc_1^{(1)}\right].
\end{equation}
If $\bsigma = -p {\bf 1} + \btau$ refers to the total stress tensor, integration of  $\bnabla\cdot\bsigma=0$ leads to the sum   of the forces, $\bf f$, on the upper and lower sheets (over a period) equal to zero, {\it i.e.}  $\mathbf{f}|_{y=y_1}+\mathbf{f}|_{y=y_2}=0$.
At leading order, the horizontal component of this relationship is
\begin{equation}
\int_0^{2\pi}\tau_{1xy}|_{y= 0}\d x=\int_0^{2\pi}\tau_{1xy}|_{y=h}\d x,
\end{equation}
which yields $\text{C}_1=0$.
We finally determine the relative velocity by insisting the sheets be force-free. Typically, for each sheet, one must sum the forces on both the inner and outer surfaces.  However, the outer problem  is force-free for all $\UDone$ \cite[]{lauga1}. The net force on the upper sheet is therefore given by
\begin{equation}
f_{1x}=-\int_0^{2\pi}a_0''(h)\d x=-2\pi\UDone /h,
\end{equation}
and hence $\UDone$ is zero, which is expected due to the $\epsilon\rightarrow-\epsilon$ symmetry of the system.
With $C_1$ and $\UDone$ equal to zero then $a_0=0$ (see \eqref{g3}) therefore we have no time-averaged flow and we get a simplified relation between stress and strain-rate in Fourier space as
\begin{equation}
\btauc_1=\frac{1+i\Detwo}{1+i\Deone}\bgamc_1.
\label{tau1c}
\end{equation}

\subsection{Second-order solution}
The second-order component of \eqref{oldroyd} is given by
\begin{align}
\left(1+\Deone\frac{\partial}{\partial t}\right)\btau_2-\left(1+\Detwo\frac{\partial}{\partial t}\right)\bgam_2=& \ \Deone\left(\bnabla\bu_1^\T\cdot\btau_1+\btau_1\cdot\bnabla\bu_1-\bu_1\cdot\bnabla\btau\right)\nonumber \\
&-\Detwo\left(\bnabla\bu_1^\T\cdot\bgam_1+\bgam_1\cdot\bnabla\bu_1-\bu_1\cdot\bnabla\bgam_1\right).
\label{oldroyd2}
\end{align}
The only part of the streamfunction, $\psi_2$, that will contribute to the force on the sheets at second order is its mean value in $x$. Using \eqref{tau1c}, the mean value of \eqref{oldroyd2} is given by
\begin{align}
\langle\btau_2\rangle-\langle\bgam_2\rangle=&\ \Re\Bigg[\frac{\Deone-\Detwo}{2(1+i\Deone)}\left(\bnabla\buc_1^{\T*}\cdot\bgamc_1+\bgamc_1\cdot\bnabla\buc_1^*-\buc_1^*\cdot\bnabla\bgamc_1\right)\Bigg],
\label{tau2}
\end{align}
where *'s indicate complex conjugates, and $\langle...\rangle$ denotes averaging over one period in $x$. The right hand side of \eqref{tau2} can then be computed using the first-order streamfunction. Upon taking the divergence and the curl of \eqref{tau2}, we obtain
\begin{equation}
\nabla^4\langle\psi_2\rangle= \frac{\Deone-\Detwo}{1+\Deone^2} \frac{\d^2}{\d y^2}G(y;h,\phi),
\label{biharmonic2}
\end{equation}
where 
\begin{align}
G(y;h,\phi)=\frac{1}{2} \bigg[&-a_1'(y) \Big(\Deone a_1''(y)+3 b_1''(y)\Big)+b_1'(y) \Big(3 a_1''(y)-\Deone b_1''(y)\Big)\nonumber \\
&+a_1(y) \Big(-2 b_1'(y)-\Deone a_1'''(y)+b_1'''(y)\Big)\nonumber\\
&+b_1(y) \Big(2 a_1'(y)-a_1'''(y)-\Deone b_1'''(y)\Big)\bigg].
\end{align}

The second-order components of the boundary conditions are
\begin{eqnarray}
\label{bc1}\boldsymbol{\nabla}\psi_2|_{y= 0} &=& -\bnabla(\UDtwo y)-\sin(x-\phi/2)\bnabla\left(\frac{\partial \psi_1}{\partial y}\right)\Big|_{y= 0},\\
\label{bc2}\boldsymbol{\nabla}\psi_2|_{y= h} &=& \bnabla(\UDtwo y)-\sin(x+\phi/2)\bnabla\left(\frac{\partial \psi_1}{\partial y}\right)\Big|_{y= h}.
\end{eqnarray}
Taking the mean value of \eqref{bc1} and \eqref{bc2}  yields
\begin{eqnarray}
\frac{\partial\langle\psi_2\rangle}{\partial x}\Big|_{y=0} &=& 0, \\ 
\frac{\partial\langle\psi_2\rangle}{\partial y}\Big|_{y=0} &=& \frac{1}{2} \left[-\UDtwo+\frac{\cos^2(\frac{\phi }{2}) (-h+\sinh h)}{h+\sinh h}+\frac{\sin^2 (\frac{\phi }{2}) (h+\sinh h)}{-h+\sinh h}\right],\\
\nonumber \\
\frac{\partial\langle\psi_2\rangle}{\partial x}\Big|_{y= h} &=& 0,\\
\frac{\partial\langle\psi_2\rangle}{\partial y}\Big|_{y= h} &=& \frac{1}{2} \left[\UDtwo+\frac{\cos^2 (\frac{\phi }{2}) (-h+\sinh h)}{h+\sinh h}+\frac{\sin^2 (\frac{\phi }{2}) (h+\sinh h)}{-h+\sinh h}\right].
\end{eqnarray}
Solving \eqref{biharmonic2} with the above boundary conditions leads to the solution
\begin{align}
\langle\psi_2\rangle=&\ \text{C}_2 y^2\left(y-\frac{3 h }{2}\right)+\frac{\UDtwo y (y-h)}{2 h}+\frac{y\cos^2 (\frac{\phi }{2}) (-h+\sinh h)}{2(h+\sinh h)}+\frac{y\sin^2 (\frac{\phi }{2}) (h+\sinh h)}{2(-h+\sinh h)}\nonumber \\
&+\frac{\Deone-\Detwo}{1+\Deone^2}  \left[\frac{y (y-2 h)}{2 h}\int Gdy|_{y=0}-\frac{y^2}{2 h}\int Gdy|_{y=h}+\int\int G dy^2\right].
\end{align}

To find  the unknown constant $\text{C}_2$ we again turn to dynamical considerations. Using integration by parts, it is straightforward to get that the force on the bottom sheet, to $O(\epsilon^2)$, is given by
\begin{equation}
f_{2x} = \int_0^{2\pi} \langle\tau_{2xy}\rangle|_{y=0}\d x,
\end{equation}
and only the  mean component of the second-order stress, $\langle\tau_{2xy}\rangle$, contributes to the net force. 
A similar relationship holds for the force on the upper sheet. 
We then proceed by obtaining  $\langle\tau_{2xy}\rangle$ from \eqref{tau2}, where $\langle\boldsymbol{\dot{\gamma}}_2\rangle=\bnabla\langle\bu_2\rangle+\bnabla\langle\bu_2\rangle^\text{T}$ and $\langle\bu_2\rangle=[\partial\langle\psi_2\rangle/\partial y,-\partial\langle\psi_2\rangle/\partial x]^\T$. Exploiting  that $\mathbf{f}|_{y=y_1}=-\mathbf{f}|_{y=y_2}$ we obtain $\text{C}_2=0$, and the net force on the upper sheet is finally given by
\begin{equation}\label{f2x}
f_{2x} =-\frac{2 \pi  \UDtwo}{h}+4\pi \left(\frac{\Deone-\Detwo}{1+\Deone^2}\right) A(h)\sin\phi,
\quad  A(h)=\frac{h \cosh h+\sinh h}{\cosh(2 h)-2 h^2-1}\cdot
\end{equation}

\section{Results}

\subsection{Synchronization}

\begin{figure}
\centerline{\includegraphics[width=.85\textwidth]{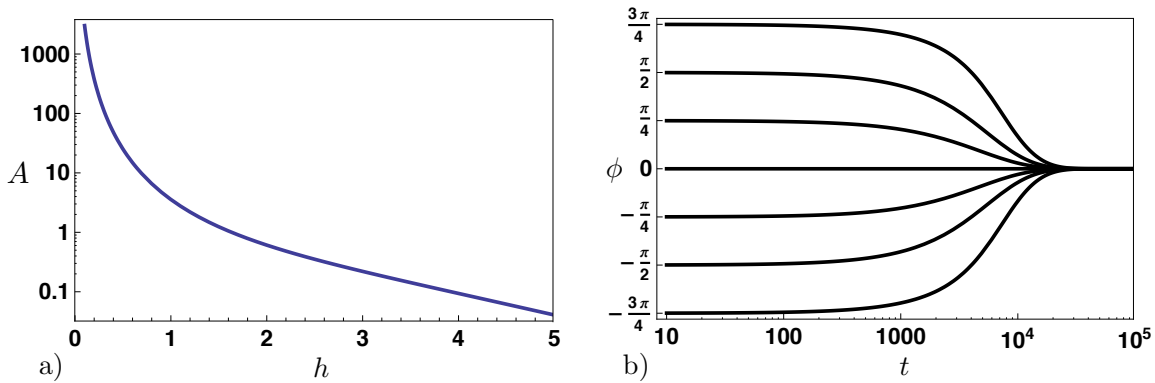}}
\caption{a) Amplitude $A$ of the phase-locking force decays exponentially with the separation distance, $h$ \eqref{f2x}. b) Time-evolution of the phase angle $\phi(t)$  from various initial conditions towards the stable in-phase conformation (\eqref{phieqn} with $\epsilon=0.1$, $h=2$, $\Deone=100$ and $\Detwo=10$).}
\label{phase}
\end{figure}

It is insightful to first consider the nature of the force which arises if the sheets are not permitted to move relative to each other but instead held with a fixed phase difference. If $\UDtwo=0$, then the force in \eqref{f2x} is zero for $\phi=0,\pm\pi$. The function $A(h)$, governing the variation in the force amplitude with mean distance $h$, is positive definite and decays exponentially with $h$, 
while becoming unbounded near $h=0$ (see Fig.~\ref{phase}a). Since $\Deone>\Detwo$, we see that the force $f_{2x}\propto \sin\phi$. This indicates that $\phi=0$ is a stable fixed point while $\phi=\pm\pi$ are unstable, and therefore we expect in-phase synchronization to occur.

We  next observe that we obtain here a nonzero force on sheets with front/back symmetric waveforms\footnote{The other symmetry, with respect to the $x$-axis, is always assumed to be true in order to enforce swimming along a straight line.}. In the case of a Newtonian fluid, this is forbidden because of kinematic reversibility, and the force is identically zero unless the front/back symmetry is broken \cite[]{elfring}. Indeed, in a Newtonian fluid, for any system with both vertical and horizontal symmetry one can reflect about both axes of symmetry then reverse the kinematics to obtain an identical conformation with the opposite force necessitating $f_x=0$ (our calculations confirm this by setting $\Deone=\Detwo=0$ in \ref{f2x}). In a viscoelastic fluid, time is no longer merely a parameter, and therefore the flow is no longer kinematically reversible, thereby permitting a nonzero force.

If instead of holding the sheets fixed, we let them move, we then have to enforce the force-free condition, and thus we obtain the relative speed $\UDtwo=hf_{2x}^s/2\pi$, where $f_{2x}^s$ is the static force incurred when $\UDtwo=0$ in \eqref{f2x}. The remarkable result is that, since the force occurs in  a viscoelastic fluid  at second order in the wave amplitude, the  phase will evolve on a time scale varying as $t\sim \epsilon^{-2}$. In a Newtonian fluid, it can be  shown that the force is always zero to second order in $\epsilon$, for any shape, and first appears at fourth order for shapes with broken front-back symmetry \cite[]{elfring}. This means that in a Newtonian fluid, at best, the phase will evolve to a phase-locked configuration on a time scale varying as $t\sim\epsilon^{-4}$. In complex fluids, synchronization is therefore seen to take place on asymptotically faster time scales than in a Newtonian fluid.

We now solve analytically for the time-evolution of the phase. Since to leading order $\dot{\phi}=-\epsilon^2\UDtwo$, we obtain a differential equation for the evolution for $\phi$ as
\begin{equation}
\frac{d\phi}{dt}=-\epsilon^2 2h \left(\frac{\Deone-\Detwo}{1+\Deone^2} \right) A(h) \sin\phi,
\label{phidot}
\end{equation}
which, for constant $h$, can be integrated to yield an analytical formula for the phase as
\begin{equation}
\phi(t)=2\tan^{-1}\left\{\tan\left(\frac{\phi_0}{2}\right) \exp\left[-\epsilon^2 2hA(h) 
\left(\frac{\Deone-\Detwo}{1+\Deone^2} \right) t\right]\right\}\cdot
\label{phieqn}
\end{equation}
Given that $\Deone > \Detwo$, we see that $\phi\sim \pm e^{-t}$ near the $\phi=0$ fixed point, meaning it is stable; however, near the $\phi=\pm\pi$ fixed points, $\phi\mp \pi  \sim   \mp e^{t}$ meaning they are unstable and hence the phase converges to $\phi=0$ for all initial conditions. The time-evolution of the phase from various initial positions, assuming a constant separation between the sheets of $h=2$, is plotted in Fig.~\ref{phase}b with $\Deone=100$, $\Detwo=10$ and $\epsilon=0.1$. All initial conformations evolve to stable in-phase synchrony. In the Newtonian case, the stability of the in-phase versus opposite-phase conformation is purely a matter of geometry, regardless of considerations of energy dissipation, and in fact two swimmers can evolve to a stable conformation which maximizes the energy dissipated \cite[]{elfring}. In contrast, in a viscoelastic fluid we find that with no asymmetry the system naturally evolves to an in-phase conformation which, as we show below, coincides with the conformation of minimal viscous dissipation.

\subsection{Energy dissipation}

The energy dissipated in the fluid between two sheets is given by integrating the dissipation density, $\btau:\bgam$, over the volume. The leading order component is given by
\begin{align}\label{dissip}
\btau_1:\bgam_1 &=\Re\left[\btauc_1\right]:\Re\left[\bgamc_1\right] =\frac{1+\Deone\Detwo}{1+\Deone^2}\bgam_1:\bgam_1-\frac{1}{2}
\frac{\Deone-\Detwo}{1+\Deone^2} 
\Im\left[\bgamc_1:\bgamc_1\right],
\end{align}
where $\Im[...]$ denotes the imaginary part. The second term in \eqref{dissip} integrates to zero over a period, thus, to leading order, the energy dissipation rate per unit depth over one period is given by
\begin{equation}\label{diss}
\dot{E}=\epsilon^2\frac{1+\Deone\Detwo}{1+\Deone^2}\int_{0}^{2\pi}\int_{y_1}^{y_2}\bgam_1:\bgam_1 \d x \d y.
\end{equation}
The result of \eqref{diss} is merely a scalar multiple of the Newtonian dissipation calculated by \cite{taylor1},  which is  minimum at the in-phase conformation, and maximum in the case of opposite-phase (and decays to zero as $h\rightarrow 0$). We see therefore that in a viscoelastic fluid the system is driven towards a state of minimum energy dissipation.

\subsection{Vertical force}
\begin{figure}
\centerline{\includegraphics[width=.85\textwidth]{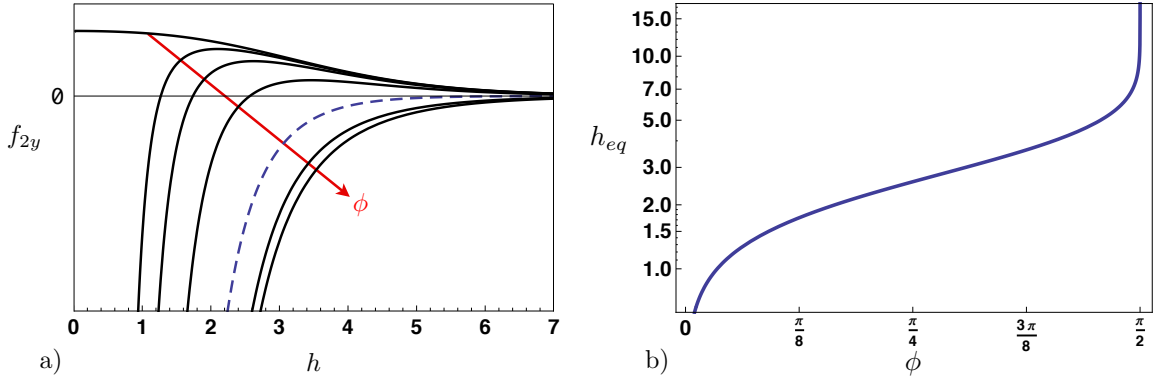}}
\caption{a) The leading-order vertical force on the lower sheet, $f_{2y}(h)$, displays a complex behavior which depends on both phase difference $\phi$ and mean separation $h$. Plotted for $\phi = \{0,\pi/16,\pi/8,\pi/4,\pi/2 \ \text{(shown dashed)},3\pi/4,\pi\}$ (arbitrary units). b) Equilibrium separation, $h_{eq}$, defined as the distance at which the vertical force is zero for a given $\phi$.}
\label{force2yb}
\end{figure}

Since the evolution of the phase depends on the separation distance $h$, it is informative to analyze the magnitude of vertical forces  between the sheets. We now proceed to compute the vertical force from the inner problem with the first and second-order streamfunctions derived here, and we use the solutions of the outer flow problem from the literature \cite[]{lauga1}. The vertical force on the bottom sheet to first order is given by
\begin{equation}
f_{1y}=\int_{0}^{2\pi}\sigma_{1_{22}}|_{y=0}-\sigma_{1_{22}}^{\rm outer}|_{y=0}\d x.
\end{equation}
Both components are individually zero, hence the force is zero. At second order, the outer flow yields no force for all $\UDtwo$ therefore the force on the bottom sheet is given by
\begin{align}
f_{2y} = \int_0^{2\pi} \left[\langle\tau_{2yy}\rangle- \int \frac{\partial \langle\tau_{2yy}\rangle}{\partial y}dy\right]\Big|_{y=0}\d x =\ 2 \pi \left(\frac{\Deone-\Detwo}{1+\Deone^2}\right) \Big[\text{B}_1(h)+\text{B}_2(h)\cos\phi \Big],
\end{align}
where
\begin{eqnarray}
\text{B}_1(h) &=& \frac{2 \left[1-\left(1+2 h^2\right) \cosh(2h)-2h \sinh(2h)\right]}{\left(\cosh(2h)-1-2 h^2\right)^2},\\
\text{B}_2(h) &=& \frac{\left(4 h^2-1\right) \cosh h+\cosh(3h)+2 h \left[3+2 h^2+\cosh(2h)\right] \sinh h}{\left(\cosh(2h)-2 h^2-1\right)^2}\cdot
\end{eqnarray}

The vertical component of the force is a cosine function in $\phi$ which is amplified by the positive-definite function $B_2$ and shifted by the negative-definite function $B_1$. Both functions become unbounded as $h\rightarrow 0$, and both tend asymptotically to zero as $h\rightarrow \infty$. In Fig.~\ref{force2yb}a we plot the vertical force as a function of the distance between the sheets, $h$, for various $\phi$ (arbitrary units). If the  phase difference is above $\pi/2$ ($\phi=\pi/2$ is shown dashed), then the sheets will be repelled from each other. However, as the sheets get closer in phase there arises a finite equilibrium separation, $h_{eq}(\phi)$, where $f_{2y}=0$. If the sheets are separated by $h<h_{eq}$, they will be repelled while if $h>h_{eq}$, they will be attracted. In Fig.~\ref{force2yb}b we plot $h_{eq}$ as a function of phase difference and we see that the equilibrium separation decreases monotonically with decreasing $\phi$ and that when the sheets are in phase the vertical force acting on them is strictly attractive. Indeed, for $\phi=0$, in the limit $h\rightarrow0$ we see that $f_{2y}=(3\pi/4)(\Deone-\Detwo)/(1+\Deone^2)$\footnote{$h$ is allowed to decrease to zero only when $\phi=0$, as otherwise the sheets would overlap.}.

\subsection{Coupled dynamics}
\begin{figure}
\centerline{\includegraphics[width=.85\textwidth]{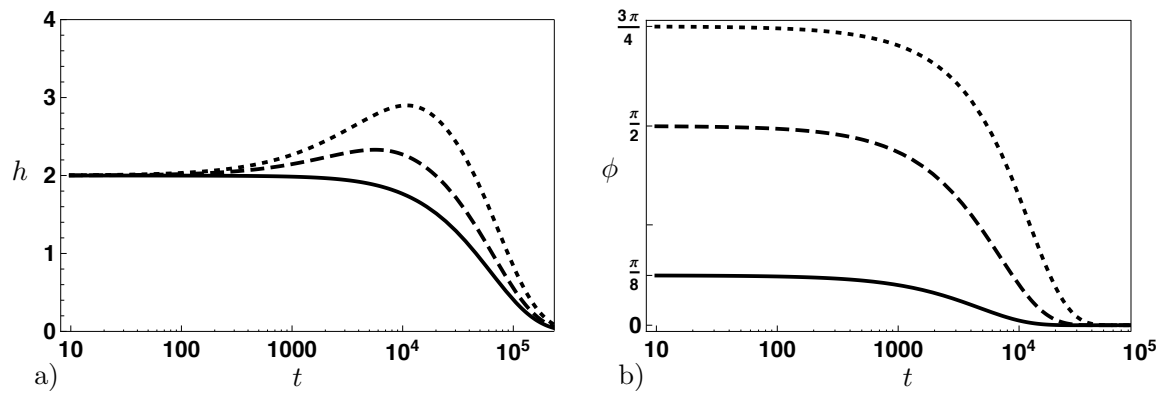}}
\caption{Coupled time evolution of the distance between the swimmers, $h(t)$ (a), and the phase difference, $\phi(t)$ (b). The initial condition is $h_0= 2$, and the mobility in $y$ is taken to be half of that in $x$:  $\phi_0=3\pi/4$ (dotted line), $\phi_0=\pi/2$ (dashed line), $\phi_0=\pi/8$ (solid line). With $\epsilon=0.1$, $\Deone=100$ and $\Detwo=10$.}
\label{phihvstime}
\end{figure}

In the idealized two-dimensional case studied here, the swimmer mobility in the vertical direction is strictly zero and hence only motion in the horizontal direction occurs. In the slender-body limit, which is the one relevant for the dynamics of  three-dimensional flagellar filaments of swimming cells, the viscous mobility in the direction perpendicular to the length of the flagellum is about half of that in the parallel direction \cite[]{kim91}. 
In order to propose a simple model for the coupled vertical/horizontal motion of the sheets, we proceed to use this ratio in our model, and  simply assume
\begin{equation}
\frac{dh}{dt}=-\frac{\epsilon^2h}{4\pi}f_{2y}^s.
\label{ymobility}
\end{equation}
Given the behavior of the vertical force, we expect the swimmers to be pushed apart slightly if their phase difference is large, then, as the phase difference decreases, to be attracted to a final synchronized conformation where the flagella are as close together as possible, as seen experimentally \cite[]{woolley}. We can numerically integrate both differential equations, \eqref{phidot} and \eqref{ymobility}, to obtain the coupled time evolution of $h(t)$, shown in Fig.~\ref{phihvstime}a, and $\phi(t)$, shown in Fig.~\ref{phihvstime}b, for an initial separation $h_0=2$. We see that for a small enough initial angle, the sheets are  monotonically attracted to each-other ($\phi_0=\pi/8$, solid line). However, for larger initial phase differences,  the sheets are initially repelled, before reaching a maximum separation, and eventually being drawn together closely. This is illustrated for $\phi_0=\pi/2$ (dashed line) and $\phi_0=3\pi/4$ (dotted line).  The time scale for the evolution of the phase angle is similar to the constant separation case, and all initial conformations converge to the stable in-phase conformation.

\section{Conclusion}

In this paper we used a two-dimensional model to  analytically address the synchronization of two swimmers in a viscoelastic (Oldroyd-B) fluid. In  Newtonian fluids, a front-back asymmetry in the swimmer's waveform is required for synchronization. In contrast,  in a viscoelastic fluid, phase-locking occurs even for swimmers displaying front/back symmetry. The two swimmers are driven into a stable in-phase conformation where a minimum of mechanical energy is dissipated, contrary to the Newtonian case where the stable conformation can be either in-phase or opposite-phase  depending only on the  waveform geometry. In addition, the evolution to a phase-locked conformation in a viscoelastic fluid occurs on asymptotically faster time scales than in a Newtonian fluid.

From a biological standpoint, the results of our model indicate that, for example, mammalian spermatozoa progressing through cervical mucus would be expected to synchronize passively, thereby reducing the work they are doing against the surrounding fluid as compared to when swimming isolated. This net energy savings could then potentially be used to increase their wave speeds, and hence swimming speed, as is observed experimentally \cite[]{woolley}. The asymptotically larger forces between swimmers might also lead to large-scale coherence in the dynamics of cell suspensions which is more pronounced in complex fluids than in Newtonian environments.


\begin{acknowledgments}
Funding by the NSF (CBET-0746285) and NSERC (PGS D3-374202) is gratefully acknowledged.
\end{acknowledgments}

\bibliography{viscoelastic}

\end{document}